\documentclass[aip,amsmath,amssymb,reprint]{revtex4-1}
\usepackage{graphicx}
\usepackage{dcolumn}
\usepackage{bm}
\usepackage[utf8]{inputenc}
\usepackage[T1]{fontenc}
\usepackage{mathptmx}
\usepackage{etoolbox}
\usepackage{multirow} 

\makeatletter
\def\@email#1#2{%
 \endgroup
 \patchcmd{\titleblock@produce}
  {\frontmatter@RRAPformat}
  {\frontmatter@RRAPformat{\produce@RRAP{*#1\href{mailto:#2}{#2}}}\frontmatter@RRAPformat}
  {}{}
}%
\makeatother

\usepackage{xcolor}
\usepackage[normalem]{ulem}

\newcommand\T{\rule{0pt}{2.6ex}}       
\newcommand\B{\rule[-1.2ex]{0pt}{0pt}} 

\begin{document}

\title{
Quantized vortex nucleation in collisions of superfluid nanoscopic helium droplets at zero temperature
}

\author{Ernesto Garc\'{\i}a-Alfonso}
\affiliation{Laboratoire Collisions, Agr\'egats, R\'eactivit\'e
  (LCAR), Universit\'e de Toulouse, CNRS, 31062, Toulouse, France}

\author{Francesco Ancilotto}
\affiliation{Dipartimento di Fisica e Astronomia ``Galileo Galilei''
and CNISM, Universit\`a di Padova, via Marzolo 8, 35122 Padova, Italy}
\affiliation{ CNR-IOM Democritos, via Bonomea, 265 - 34136 Trieste, Italy }

\author{Manuel Barranco}
\affiliation{Departament FQA, Facultat de F\'{\i}sica,
Universitat de Barcelona, Av.\ Diagonal 645,
08028 Barcelona, Spain.}
\affiliation{Institute of Nanoscience and Nanotechnology (IN2UB),
Universitat de Barcelona, Barcelona, Spain.}

\author{Mart\'{\i} Pi}
\affiliation{Departament FQA, Facultat de F\'{\i}sica,
Universitat de Barcelona, Av.\ Diagonal 645,
08028 Barcelona, Spain.}
\affiliation{Institute of Nanoscience and Nanotechnology (IN2UB),
Universitat de Barcelona, Barcelona, Spain.}

\author{Nadine Halberstadt}
\affiliation{Laboratoire Collisions, Agr\'egats, R\'eactivit\'e
  (LCAR), Universit\'e de Toulouse, CNRS, 31062, Toulouse, France}

\begin{abstract}
We address the collision of two superfluid $^4$He droplets 
at non-zero initial relative velocities and impact parameters
within the framework of liquid $^4$He time-dependent density functional theory at zero temperature. 
In spite of the small size of these droplets (1000 He atoms in the merged droplet) 
imposed by computational limitations, we have found that quantized vortices may be readily nucleated for reasonable collision parameters. 
At variance with head-on collisions, where only vortex rings are produced, collisions with non-zero impact parameter
produce linear vortices which are nucleated at indentations appearing on the surface of the deformed merged droplet.
Whereas for equal-size droplets vortices are produced in pairs,
an odd number of vortices can appear  when the colliding droplet sizes are different.
In all cases vortices coexist  with surface capillary waves. 
The possibility for collisions to be at the origin of vortex 
nucleation in experiments involving very large droplets is discussed.
An additional surprising result is the observation of the drops 
coalescence even for grazing and distal collisions at relative velocities as high as 80~m/s and 40~m/s, respectively,
induced by the long-range Van der Waals attraction between the droplets.

\end{abstract}

\date{\today}

\maketitle

\section{Introduction}
Superfluid helium droplets are routinely produced in beams  obtained  by expanding the high purity gas or liquid through a nozzle  into vacuum.
  The temperature $T_0$ and pressure $P_0$ 
values at the  source chamber and the characteristics of the nozzle determine the appearance of the jet 
 and the size and velocity of the droplets.\cite{Toe04} Once formed, drops cool down by evaporative cooling, eventually  becoming superfluid. 
 The activity in the field has been  comprehensively presented in a recent monograph.\cite{Sle22}

The study of vortices in helium droplets has been a 
subject of continuous interest since they were first detected in droplets made of $\sim 10^{8-11}$ atoms,\cite{Gom14,Ges19} 
hereafter referred to as ``very large droplets'' (VLD).
VLD are believed  to acquire angular momentum as they pass through  the nozzle.
As a result of the superfluid transition, most angular momentum deposited in the droplet is stored in nucleated quantized vortices,
 while some remains as surface capillary waves in the deformed droplets and some is taken away by evaporated He atoms. 
The morphology of these VLD has been
addressed in detail,\cite{Lan18} and the coexistence of quantized vortices and capillary waves has been established.\cite{Oco20,Pi21}  

Capture of impurities by droplets may also lead to vortex nucleation. Indeed, it has been shown that 
impurity capture by droplets made of $N=1000$ atoms produces vortex rings and vortex loops.\cite{Mat14,Lea14,Cop17} However,
detecting vortices in these small droplets is a challenge. Methods based on studying the absorption spectrum of atomic impurities
attached to the vortex cores have been proposed\cite{Clo98,Her08,Gar20} but so far vortices in small droplets
have eluded detection.

In this work we concentrate on the study of an alternative vortex formation mechanism, namely droplet-droplet collisions at non-zero 
impact parameter. Experiments on He drops collisions, although feasible in principle, have not been carried out; they would be
technically challenging and require rather expensive cryogenic cooling. At variance, 
molecular-beam scattering experiments where a beam of He droplets interacted with a secondary beam of  Ar or Kr atoms  
have been performed to determine the appearance of $^4$He and $^3$He droplets made of $O(10^3-10^4)$ atoms.\cite{Har98,Har01}
Let us mention that coalescence experiments of helium droplets magnetically levitated   have been carried out.
Using a static magnetic field, drops of less than 1 cm radius at a temperature of 0.7 K were confined and made to collide at velocities as small as 
a few cm/s.\cite{Vic00} 

Recently,  experimental activity has been conducted on the fragmentation of thin liquid helium jets into vacuum.\cite{Tan20,Kol22}
It has been found that under suitable conditions, equidistant droplets with almost uniform size are produced from the breakup of the jet, and
that sometimes these drops coalesce downstream.\cite{Tan20,Kol22,Ulm23}
These  droplet collisions can occur because of the spread of droplet velocities inside 
the jet,\cite{Gri03} which although small can be the source of non-zero relative velocity and impact parameter.  

Our goal is to describe binary collisions 
of zero temperature superfluid $^4$He droplets within the 
$^4$He density functional (He-DFT) approach.\cite{Dal95,Bar06,Anc17,dft-guide}
This approach is similar, in the superfluid $^4$He phase, to the Gross-Pitaevskii (GP) approach  which
has successfully been applied to the description of cold gases in the superfluid Bose-Einstein condensate phase, in particular in
the study of quantized vortices.\cite{Pit16,Bar16,Tsu09}

In a recent work, some of us have addressed the coalescence of superfluid $^4$He  droplets,\cite{Esc19}  initially at rest, which were drawn together by their mutual Van der Waals (vdW) long-range attraction. 
 The merging of vortex-free helium droplets has unveiled the appearance of vortex-antivortex ring pairs nucleated at the droplet 
 surface, that either wrap around the coalesced droplet or penetrated into it, eventually annihilating each other 
 yielding an intense roton burst. This work has been later extended to the case of vortex-hosting droplets.\cite{Esc22}
 To our knowledge, no other description of superfluid (i.e. inviscid and irrotational) $^4$He droplet collisions is available in the literature.
We want to mention the existence of theoretical and experimental studies on head-on collisions of ``quantum droplets'' 
made of a very low temperature  gas of $^{39}$K atoms in two different hyperfine states  constituting a 
superfluid Bose mixture,\cite{Fer19,Cik21} 
which bears some similarities with the problem of $^4$He droplets collisions.

Binary collisions of droplets made of viscid fluids occur in, e.g.,  raindrop formation or spray processes. 
Besides initial velocity and impact parameter, the collision 
outcome depends on the rheological properties of the droplets:  droplet bouncing, droplet coalescence and drop stretching separation have been found with increasing Weber number. 
 It is worth mentioning that
$^3$He droplets collisions were described long ago in the Vlasov dynamics.\cite{Gui95} These drops were found to bear collision 
properties that, on the one hand, are common to classical mesoscopic systems, like e.g., mercury drops\cite{Men93,Men97}  
and, on the other hand, are common to heavy-ion reactions, like fusion-fission and deep-inelastic processes.\cite{Ngo86}

Classically, binary collisions are addressed by solving the Navier-Stokes (NS) and continuity equations subject to appropriate 
boundary conditions, see e.g. Ref. \onlinecite{Nik09} and references therein.
It is naturally assumed that the solution of the NS equation for small enough viscosities should be 
nearly indistinguishable from the inviscid limit.\cite{Ant19,Hoe13} However,  as emphasized in Ref. \onlinecite{Anc23}, neither
 time-dependent GP nor He-DFT equations appropriate for superfluids 
reduce to the zero-viscosity limit of the NS equation (Euler equation) for a barotropic fluid in irrotational flow.\cite{Bar16} 
Indeed, in the superfluid case an extra term appears involving the gradient of the so-called quantum pressure $Q$
\begin{equation}
Q=\frac{\hbar^2}{2m_4} \, \frac{\nabla^2 \rho^{1/2}}{\rho^{1/2}} \, ,
\end{equation}
where $m_4$ is the mass of the $^4$He atom and $\rho$ is the atom density. This term  plays
a crucial role when the density is highly inhomogeneous, as is the case near the core of a quantized vortex  for instance.
Quantum pressure  is a key ingredient naturally included in our time-dependent He-DFT 
approach.

Helium density functional and time-dependent density functional (He-TDDFT) methods
have proven to 
be very powerful tools to study the properties and dynamics of superfluid $^4$He droplets. 
Within the He-DFT approach, the finite range of 
the helium-helium van der Waals (vdW) interaction is explicitly  incorporated in the
simulations. As a consequence, the liquid-vacuum interface
has a non-zero surface width, which is important in the 
description of nanoscopic $^4$He droplets like those studied in the present work.
The finite compressibility of the fluid is taken into account, and therefore possible
density excitations (ripplons, phonons and rotons) are naturally reproduced. 
The possibility for atom evaporation 
from the $^4$He sample during the real-time dynamics is also included.\cite{Gar22}

This work is organized as follows. In Sect. II we briefly present the He-DFT approach.
In Sect III we discuss the results obtained for the collision dynamics.
Due to the computational burden associated with fully three-dimensional
He-DFT simulations, 
we only address a few illustrative cases corresponding to selected values of the initial droplets velocity and impact parameter.
A summary with some concluding remarks is presented in Sect. IV. 
In complement to the main text,  the supplementary material  provides movies of
the real-time dynamics of the $^4$He droplet collisions addressed in this
work. This multimedia material  constitutes an important part of this work, since  
it helps capture physical details which would otherwise escape the written account.

\section{Method}
  
  To describe the droplet-droplet collisions
  we have applied the $^4$He density functional (DFT) and time-dependent density functional (TDDFT) methods thoroughly 
  described in Refs.  \onlinecite{Anc17,dft-guide}.     
Let us briefly recall that within DFT, the energy of the droplet is written as a functional of the atom density $\rho({\mathbf r})$ as
\begin{equation}
E[\rho] = T[\rho] + E_c[\rho] =
\frac{\hbar^2}{2m_4} \int d {\mathbf r} |\nabla \Psi({\mathbf r})|^2 +  \int d{\mathbf r} \,{\cal E}_c[\rho]
\label{eq2}
\end{equation}
where the first term  is the kinetic energy with $\rho({\mathbf r})= |\Psi({\mathbf r})|^2$
and  the functional ${\cal E}_c$ contains the interaction term (in the
Hartree approximation) and additional terms which describe non-local correlation effects.\cite{Anc05}
  
The droplet equilibrium configuration is obtained by solving the Euler-Lagrange equation resulting 
from the functional variation of Eq.\ (\ref{eq2}),
\begin{equation}
\left\{-\frac{\hbar^2}{2m_4} \nabla^2 + \frac{\delta {\cal E}_c}{\delta \rho}  \right\}\Psi({\mathbf r}) 
 \equiv {\cal H}[\rho] \,\Psi({\mathbf r})  = \mu \Psi({\mathbf r})
\;,
\label{eq3}
\end{equation}
where $\mu$ is the $^4$He chemical potential corresponding to the number of He atoms in the droplet, 
$N = \int d{\bf r}|\Psi({\bf r})|^2$. 

To prepare the collision, we have first calculated the equilibrium structure of a $^4$He$_{500}$ droplet.
We have found it convenient to
obtain the structure of each single droplet inside the larger calculation box where  the dynamics will be carried out, placing their centers
of mass so that their dividing surfaces (loci where the helium density equals half the liquid density value,
$R=r_0 N^{1/3}$, with $r_0= 2.22$ \AA{}) are 8 \AA{} apart and 
the impact parameter equals the chosen value. This yields two equal density
profiles centered at
different points of the calculation box, $\rho_1(\mathbf{r})$ and $\rho_2(\mathbf{r})$.

Next,  we build an effective wave function giving  the droplets opposite velocities in the $z$ direction as follows:
\begin{equation}
\Psi(\mathbf{r}, t=0) = e^{-i k z} \sqrt{\rho_1(\mathbf{r})}+e^{i k z} \sqrt{\rho_2(\mathbf{r})} \; ,
\label{eq4}
\end{equation}
where the wave number $k$ is related to the droplet velocity  $v$ as $v = \hbar k/m_4$. The TDDFT equation 
\begin{equation}
i\hbar \frac{\partial}{\partial t}\Psi(\mathbf{r}, t) = {\cal H}[\rho] \, \Psi(\mathbf{r}, t) 
\label{eq5}
\end{equation}
is solved taking Eq. (\ref{eq4}) as the starting effective wave function. 

We have set $(y,z)$ as the reaction plane and the $z$ axis as  direction of incidence.
The angular momentum, written in units of $\hbar$ thorough this paper, is calculated as
\begin{equation}
L(t) = - i \int d \mathbf{r} \, \Psi^*(\mathbf{r}, t) \left( y \frac{\partial}{\partial z}- z \frac{\partial}{\partial y} \right)\Psi(\mathbf{r}, t) \, ,
\label{eq6}
\end{equation}
where $ -i (y\partial/\partial z- z \partial/\partial y = \hat{L}_x$ is the angular momentum operator in the $x$-direction.

In practice, Eqs. (\ref{eq3}) and (\ref{eq5}) have been solved using the $^4$He-DFT-BCN-TLS 
computing package,\cite{Pi17} see  Refs.~\onlinecite{Anc17} and \onlinecite{dft-guide} and references therein for details.
We work in cartesian coordinates, with $\rho_i(\mathbf{r})$ and 
$\Psi(\mathbf{r})$ defined at the nodes of a 3D grid inside a
calculation box large enough to accommodate the droplets in such a way 
that the He density is sensibly zero at the box surface.
Periodic boundary conditions are imposed so that the convolutions involved in the DFT mean field ${\cal H}[\rho]$ can be carried out  using Fast Fourier Transform.\cite{Fri05} 
The differential operators in ${\cal H}[\rho]$  are approximated by 13-point formulas.
All the simulation grids had $288$ points equally spaced by $0.4$~\AA\  in each direction, except for the head-on collision for which the grid had to be extended along the incidence axis ($z$) to $576$ points.

The TDDFT equation is
solved using Hamming's predictor-modifier-corrector method\cite{Ral60}
initiated by  a fourth-order Runge-Kutta-Gill method,\cite{Ral60} with 
a time step of $0.1$~fs.
During the time evolution some helium may evaporate from the droplets, eventually reaching the cell boundary. 
To prevent  this material from reentering the cell due to the imposed periodic boundary conditions, we include
an absorption buffer of $2$~\AA\ inside the calculation box\cite{Mat11,dft-guide} in each direction.
This particle --and thus angular momentum and energy-- leaking 
is obviously physical. Space and time steps have been chosen to keep energy and angular momentum well conserved 
in the absence of atom evaporation.

\section{Results}

 Our main goal is  to investigate whether vortices could be nucleated during a droplet-droplet collision for reasonable values of their initial velocity and impact parameter.
Let us first obtain a crude estimate of the critical impact parameter $b_{cr}$ leading to vortex nucleation.  
As stated in the Introduction, we assume that the droplet collision may result
from the finite velocity dispersion in the droplet beam.


If  $\Delta v$ is the velocity spreading in the jet system of reference, which moves with a velocity $v_j$ with respect to the laboratory system, the maximum relative velocity is $2 \Delta v$.
The angular momentum $L$ created in the merged droplet is given by  $L \hbar = b N m_4 \Delta v $, where $N$ is the number of helium atoms in each colliding droplet.
For a vortex line along the diameter of the coalesced spherical droplet one has $L=2N$, which yields the critical impact parameter
\begin{equation}
b_{cr} = 2 \frac{\hbar}{m_4} \frac{1}{\Delta v}
\label{eq1}
\end{equation}

Let us take as an example the  velocity range $29 \leq v_j \leq 310$ m/s   explored in the 
experiments of  Kolatzki et al.,\cite{Kol22} 
in which droplet beams are obtained by fragmentation of a thin liquid helium jets into vacuum.\cite{Tan20,Kol22}
Under these experimental conditions, $\Delta v/v \sim 0.01$,\cite{Gri03} i.e., $0.3 \leq \Delta v \leq 3$ m/s,
 hence $106.7 \leq b_{cr} \leq 1067$ \AA{}. 
 For a grazing collision, $b_{cr} = 2R$ hence $53.4 \leq R \leq 534$ \AA{}, thus 
giving $ 1.4 \times 10^4 \leq N \leq 1.4 \times 10^7$. 
 
As indicated above, this is only a crude estimate.
 Even if enough angular momentum is available from the start, a vortex will not necessarily be nucleated since
   part of the angular momentum will be stored in capillary waves.\cite{Pi21}  
 This is all the more true since the merged droplet will be deformed for quite some time.
 Also, it is not obvious a priori if a grazing collision can lead to droplet coalescence.
On the other hand, less angular momentum is required to nucleate  a non-centered vortex line.\cite{Bau95,Leh03}
 
 The previous estimate makes it clear that a realistic simulation
 of the collision process between droplets arising from jet breaking in usual experimental conditions 
 is beyond the  TDDFT capabilities due to the large size of the involved droplets. 
 On the other hand, the whole collision process can be simulated in detail for smaller droplets.
 Experimentally, they are obtained in a different expansion regime, 
called regime 1 or supercritical in the recent review by Toennies,\cite{Toe22} 
in which droplets are formed by gas condensation.
 Depending on experimental conditions, their size can vary from several atoms up to about 10000 atoms.
 For instance in a $5$~$\mu$m diameter nozzle at $P_0=80$~bar and $T_0=24$~K, the maximum of the log-normal size distribution has been measured to be 
 $1930$.\cite{Lew93}
 In these conditions the beam velocity is $480$~m/s\cite{Lew93} and the velocity spread is $\Delta v/v\approx 2$\%.\cite{Har98}
 These conditions would give $\Delta v\approx 10$~m/s and $b_{cr}=31.8$~\AA.
 For a grazing collision this would correspond to a droplet radius of about $16$~\AA, slightly smaller than the radius of a 500-atom droplets ($17.6$~\AA) investigated in our work.
 Higher nozzle temperatures lead to log-normal size distributions peaking at lower sizes.
 
  We address here the collision, at non-zero relative velocity and  impact parameter,
 of two $^4$He$_{500}$ droplets of radius $R=r_0 N^{1/3}$ with $r_0= 2.22$ \AA{}, i.e., $R=17.6$ \AA{}. 
From Eq. (\ref{eq1}), for a not so grazing collision with  $b =3R/2$, $\Delta v=12$ m/s whereas for a more
  central collision with $b=R$, $\Delta v=$ 18 m/s. Since part of the angular momentum will go into  capillary waves or will be taken
  away by atom evaporation, in our study we have also considered two larger values for  $\Delta v$, namely 20 and 40 m/s.
   Specifically, we have chosen as cases of study the following combinations of droplet velocity $v$ and impact 
  parameter $b$:
  
  \begin{itemize}
  \item
 $b=0$, $v=$ 40 m/s (head-on collision).
 \item
 $b=3R/2$, $v=$10, 20, and 40 m/s.
 \item
 $b=2R$, $v=$20 and 40 m/s (grazing collision).
 \item
 $b=5R/2$, $v=$20 m/s (distal collision, $b> 2R$).
 \end{itemize}
       
These selected values allow for comparing the results at a given impact parameter as a function of the initial velocity, and the other way around.
Since droplets are equal, the relative velocity in the collisions is $v_{rel}=2v$. 
We have also studied a non-symmetric case of two droplets, one of 300 atoms and the other of 700 atoms.
 
 \subsection{Head-on collision at $v= $ 40 m/s}
\begin{figure}[t]
 \hspace*{-0.5cm}
\includegraphics[width=1.12\linewidth,clip=true]{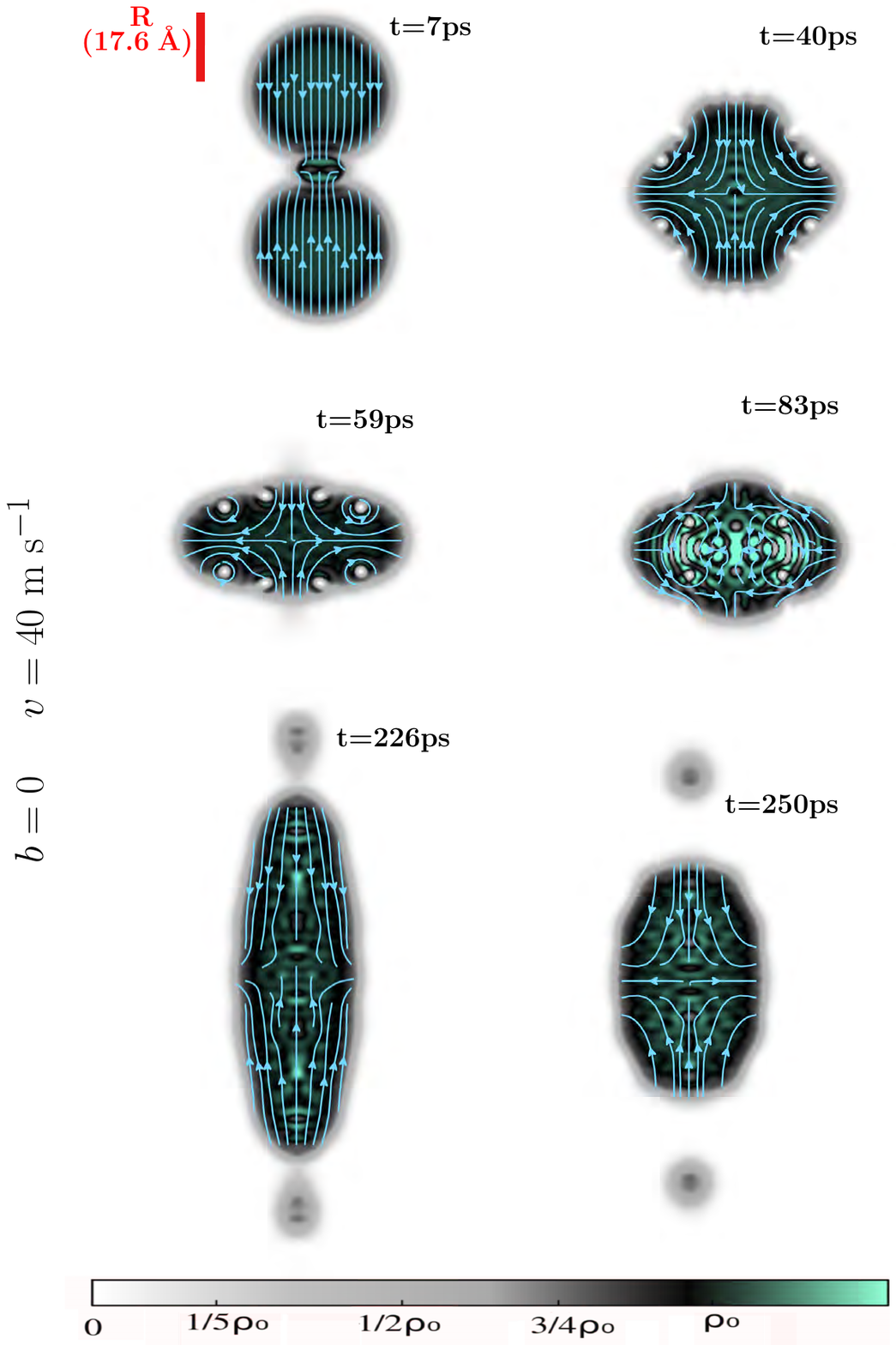}
\caption{\label{fig:b0-v04-multiplots} 
Snapshots of 2-dimension cuts in the $(y,z)$ collision plane for impact parameter $b=0$ (``head-on collision'') and $v=40$~m/s.
$\rho_0$ in the density scale is the bulk superfluid helium density, $\rho_0=0.0218$~\AA$^{-3}$.
Superimposed arrows represent the superflow current.}
\end{figure}

 This is a zero-angular momentum collision.
 Figure~\ref{fig:b0-v04-multiplots} 
 shows snapshots  of  2D density cuts in the $(y,z)$  collision plane  during the real-time dynamics.
 Superimposed to the density we have plotted the superflow current. This format is common to all 2D density figures in this paper.
  Upon droplet contact, due to the fairly large relative velocity, a density bulge develops at the collision region (frame at 7 ps)
 which expands laterally because
 of the large incompressibility of helium. This bulge is absent in the simulation of two droplets drawn 
 against each other\cite{Esc19} only by the vdW attraction because of the smaller velocity involved in that process. 
 
 One may see the nucleation of vortex rings at  surface indentations (frame at 40 ps). Due to the symmetry of the process, 
 vortices appear in pairs of rings-antirings. As in Ref. \onlinecite{Esc19}, vortex rings/antirings are also nucleated at the density protrusions 
 symmetrically placed along the collision direction (frame at 59 ps). These ring pairs eventually collide and annihilate, producing a
 roton burst (frame at $t=$ 83 ps). 
As discussed in the following, the density waves produced by the 
rings annihilation induce He atom evaporation as they reach the droplet surface. 
After the fusion, the merged droplet in the figure undergoes wide amplitude oscillations. 
Interestingly, two satellite droplets 
appear (frame at 226 ps) 
 which eventually  detach from the fused droplet (frame at 250~ps). Any residual friction/viscosity remaining in the system might hinder
 this process. Density oscillations are also expected to be damped for the same reason. 
 
\begin{figure}[t]
 \hspace*{-0.cm}
\includegraphics[width=0.85\linewidth,clip=true,angle=270]{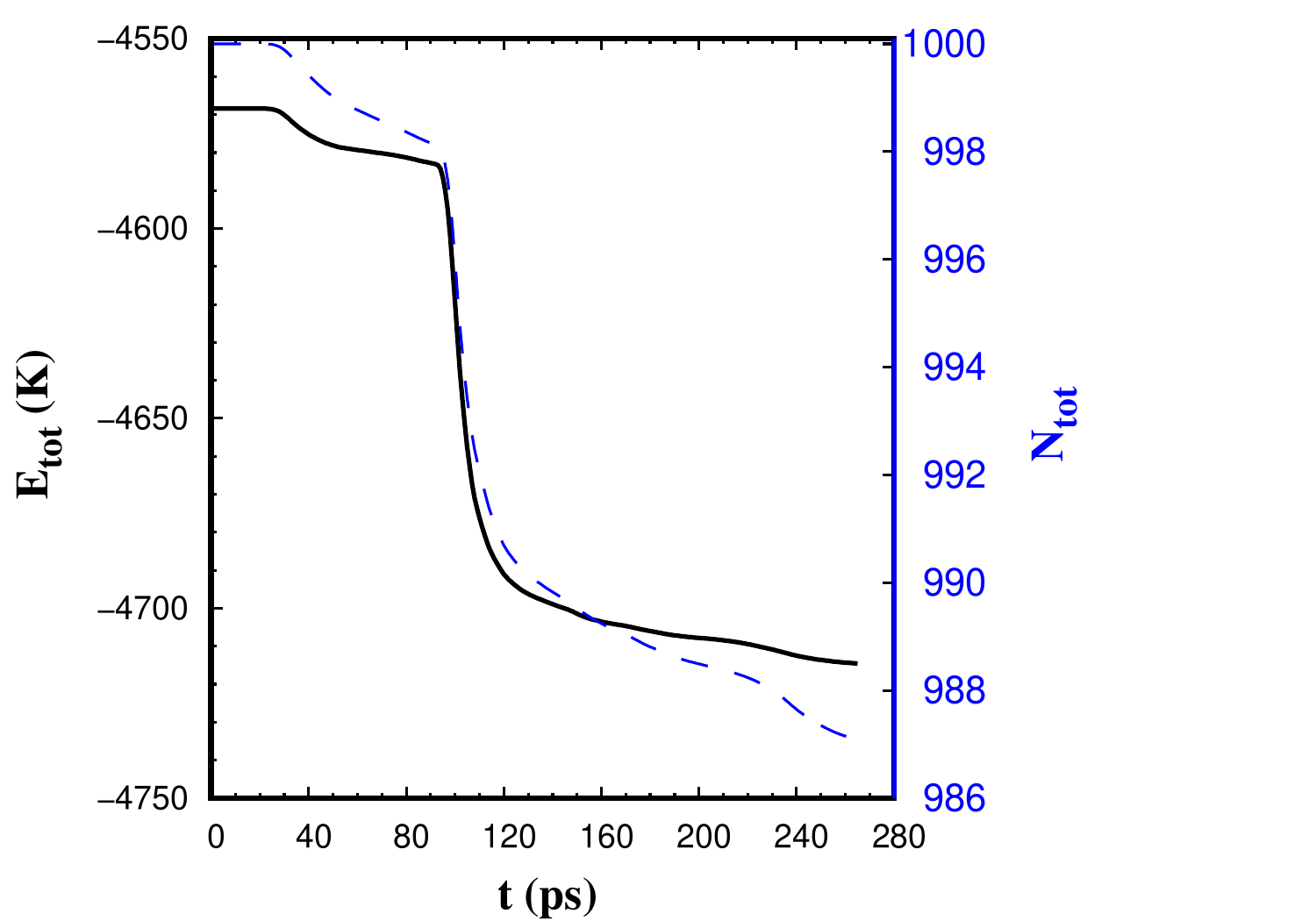}
\caption{\label{fig:b0-v04-EN} 
Time evolution of the total energy $E_\text{tot}$  (black line, left vertical axis) and number of helium atoms $N_\text{tot}$ (dashed blue line, right vertical axis)  
for He$_{500}$ + He$_{500}$ collision with impact parameter $b=0$  (``head-on collision'') and $v=40$~m/s.}
\end{figure}

 Notice that atom evaporation, which also
 contributes to the damping, is naturally occurring in our simulations. 
 Figure~\ref{fig:b0-v04-EN} 
 shows the time evolution of the energy and atom number in the system.
 The roton burst observed in the   $t=$ 83~ps snapshot induces strong helium atom evaporation: $\sim 8$~atoms in $\sim 20$~ps, 
 dissipating $\sim110$~K (about 14~K/atom).
 This is followed by slower atom evaporation and energy dissipation.
 During the time elapsed by the simulation (265 ps), 13 He atoms are emitted taking away an average energy of about 11~K/atom.

\subsection{Collisions with impact parameter $b=3R/2$} 
\subsubsection{$v=$ 10 m/s collision} 
 
\begin{figure}[t]
 \hspace*{-0.5cm}
\includegraphics[width=1.1\linewidth,clip=true]{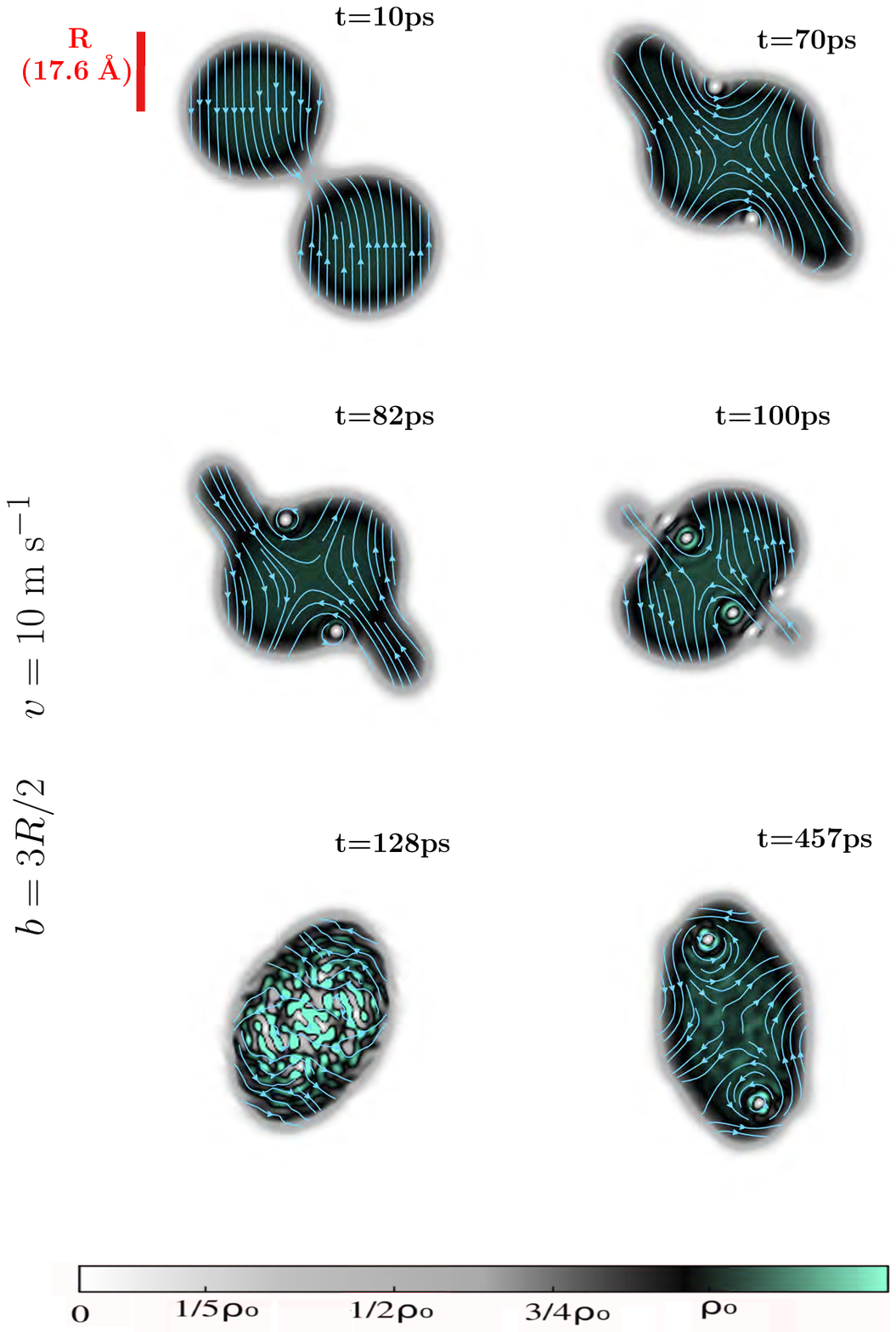}
\caption{\label{fig:b3R2-v01-multiplots} 
Snapshots of 2-dimension cuts in the $(y,z)$ collision plane for impact parameter $b=3R/2$ and $v=10$~m/s.
}
\end{figure}
 
\begin{figure}[t]
 \hspace*{-0.cm}
\includegraphics[width=0.85\linewidth,angle=270,clip=true]{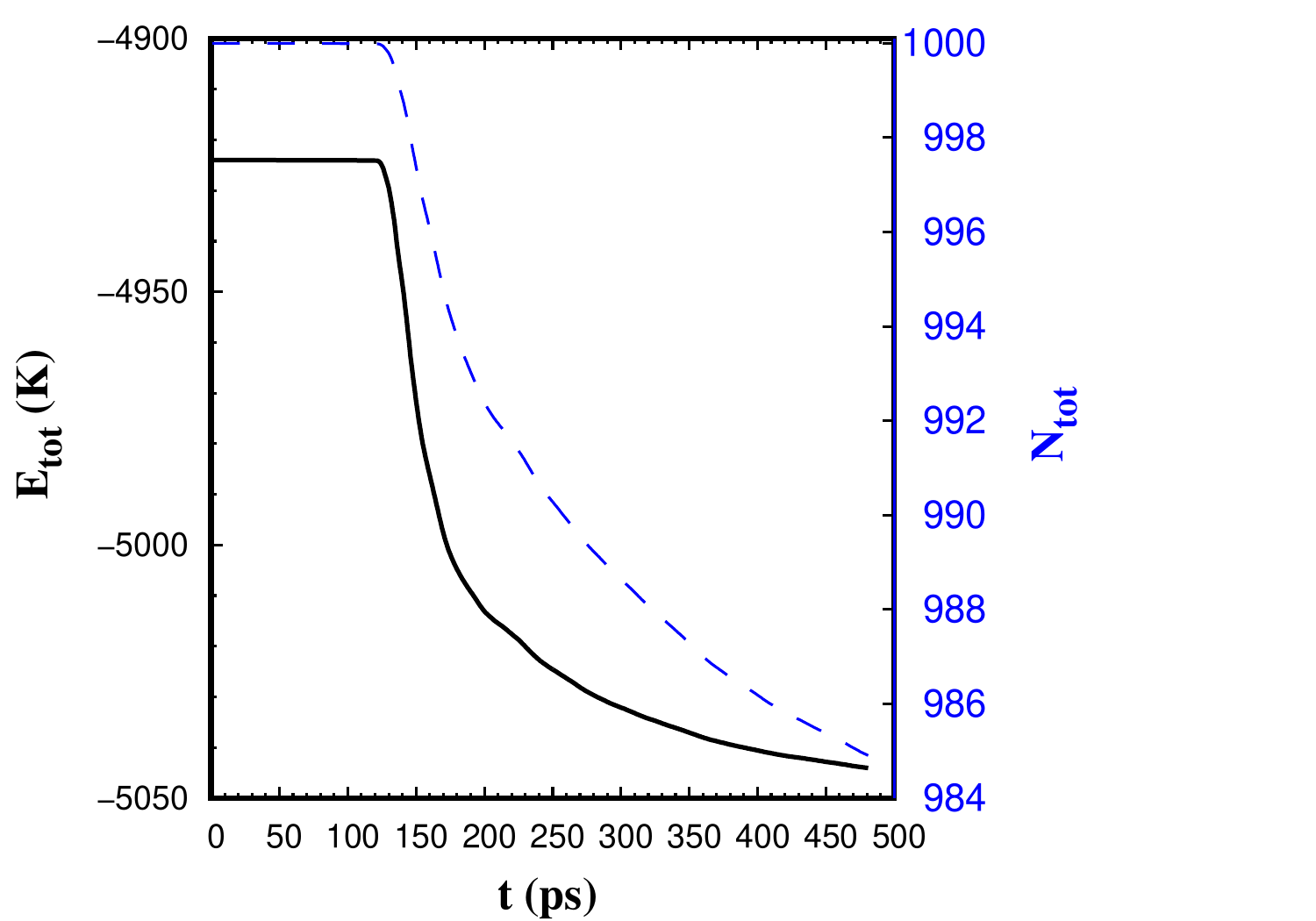}
\caption{\label{fig:b3R2-v01-EN} 
Time evolution of the total energy (black line, left vertical axis) and number of helium atoms $N_\text{tot}$ (dashed blue line, right vertical axis)  for He$_{500}$ + He$_{500}$ collision with impact parameter $b=3R/2$ and $v=10$~m/s.}
\end{figure}

The angular momentum involved in this collision is $L=825$.
Figure~\ref{fig:b3R2-v01-multiplots} 
shows snapshots of the 2D density during the collision process which display several interesting features.
A low-density bridge appears between the droplets before touching due to the long range attractive vdW mutual interaction, here exemplified by the frame at
10 ps. Interestingly, in spite of the small velocity, quantized vortices are nucleated at surface indentations appearing at the droplets
contacting region (frame at 70 ps). 
It is worth noting that a linear vortex can be nucleated even though $L=825$ is smaller than the the number of atoms in the merged droplet $N_\text{He}=1000$ :
The formula $L\hbar=N_\text{He}\hbar$ is strictly valid only  for a linear vortex along the symmetry axis of an axisymmetric droplet, 
$L$ being smaller for vortex lines displaced off the symmetry axis.\cite{Bau95,Leh03}%

Here these vortices appear in pairs because of the symmetry of the system; they are vortex lines (not rings)
of equal circulation, constituting a vortex dimer.\cite{Esc22} 
Surface protrusions appear as in the head-on collision case 
(frame at 82 ps), and their collapse nucleates a pair of vortex-antivortex rings (frame at 100 ps). The interaction of the vortex dimer
with the pair of vortex rings inside the small volume of the fused droplet causes the annihilation of the ring pair and the appearance of a roton
burst, leaving the droplet into a turbulent state (frame at 128 ps). Eventually, the fused droplet pacifies yielding a  droplet in apparent rotation alongside with
the vortex dimer inside it, as shown in the $t=457$ ps frame.

We thus see that vortices are readily nucleated in the course of the collision even for moderate values of the relative velocity and impact parameter. At long times, 
the merged droplet ``rotates'' adopting an ellipsoidal-like shape, inside which the vortex dimer moves. We shall estimate later (section~\ref{ssec:cap-vx-L})
how angular momentum  is shared between capillary waves, responsible for  the apparent rotation of the droplet, and the vortex dimer.

Figure~\ref{fig:b3R2-v01-EN} 
shows the time evolution of the energy and atom number in the system.
Atom evaporation starts around $130$~ps, when turbulence sets in, then it gradually slows down.
During the first 480 ps, about 15 He atoms are evaporated, taking away an average energy of $8$~K per atom and an angular momentum 
of  about 2.3 units per  atom.
Note that the initial excess energy with respect to a vortex free 1000-He atom droplet is 476~K in this case, so that the merged droplet still contains a significant amount of internal energy at the end of the simulation, even taking into account the additional energy contained in a vortex-hosting droplet.
Unfortunately it is not possible to continue the simulation to much longer times.

The results for all the collisions studied in this work are collected in Table~\ref{tab:Summary_coalescence}.

\begin{table}[h!]
\renewcommand{\arraystretch}{1.0}
\setlength{\tabcolsep}{2mm}
    \hspace{0.cm}\begin{tabular}{c c r@{\hspace*{0.3cm}} | c r @{\hspace*{0.3cm}} r r} \cline{1-7}  
        \hline
        \hline
     \multirow{2}*{$b$}     &$v$  &  \multirow{2}*{$L_0$}  &  $t_f$ &  \multirow{2}*{$L_f$} &  \multirow{2}*{$n$} & $\Delta E/n$   \T \\  
                                       & (m/s)   &   &   (ps) &  &  &  (K)   \T \\  
\hline
       $0$    & 40    & 0     & 265   & 0        & 12.9   & $11.3$  \T\\
    $ 3R/2$ & 10   & 825  & 480  & 791    & 15.0   & $8.0\,\,$ \\
                 & 20   &1650 & 359  & 1549   & 10.0   & $9.2\,\,$  \\ 
                 & 40   &3300 & 362  & 3062   & 15.7   & $6.2\,\,$  \\   
    $2R$     & 20   &2200 & 140  & 2166   &  1.3   & $16.7\,\,$  \\ 
                  & 40   &4400 & 469  & 4280   &   5.5   & $5.1\,\,$   \\  
    $5R/2$   & 20   &2750  & 691  & 2683   &  3.0   & $3.6\,\,$  \B \\  
\hline
 asym.                & $v_1$=28 & \multirow{2}*{1650} & \multirow{2}*{306} & \multirow{2}*{1533} &  \multirow{2}*{7.3}   & \multirow{2}*{$8.6$} \T \\  
 $(25/21) (3R/2)$ & $v_2$=12 &  &  &  &    &  \B\\  
        \hline
        \hline
    \end{tabular}
        \caption{Summary of the results for He$_{500}$ + He$_{500}$ collisions, except for the last line, labeled \emph{asym.} 
        which refers to He$_{300}$ + He$_{700}$ collision.
    The first 3 columns correspond to initial conditions: impact parameter $b$  ($R=17.6$~\AA, the 500-droplet radius);
    initial droplet velocity $v$ in m/s (in the case of the asymmetric collision the  He$_{300}$ velocity is $v_1=28$~m/s and the He$_{700}$ velocity is $v_2=12$~m/s) ; 
    resulting initial angular momentum $L_0$.
    $t_f$ is the total duration of the simulation; 
    $L_f$ is the angular momentum at the end of the simulation; 
    $n$ is the number of evaporated helium atoms;
    $\Delta E/n$ is the average energy loss per evaporated atom.
    \label{tab:Summary_coalescence}}
\end{table}

\subsubsection{$v=$ 20 m/s and  40 m/s collisions } 
 
\begin{figure}[t]
 \hspace*{-0.5cm}
\includegraphics[width=0.85\linewidth,angle=270,clip=true]{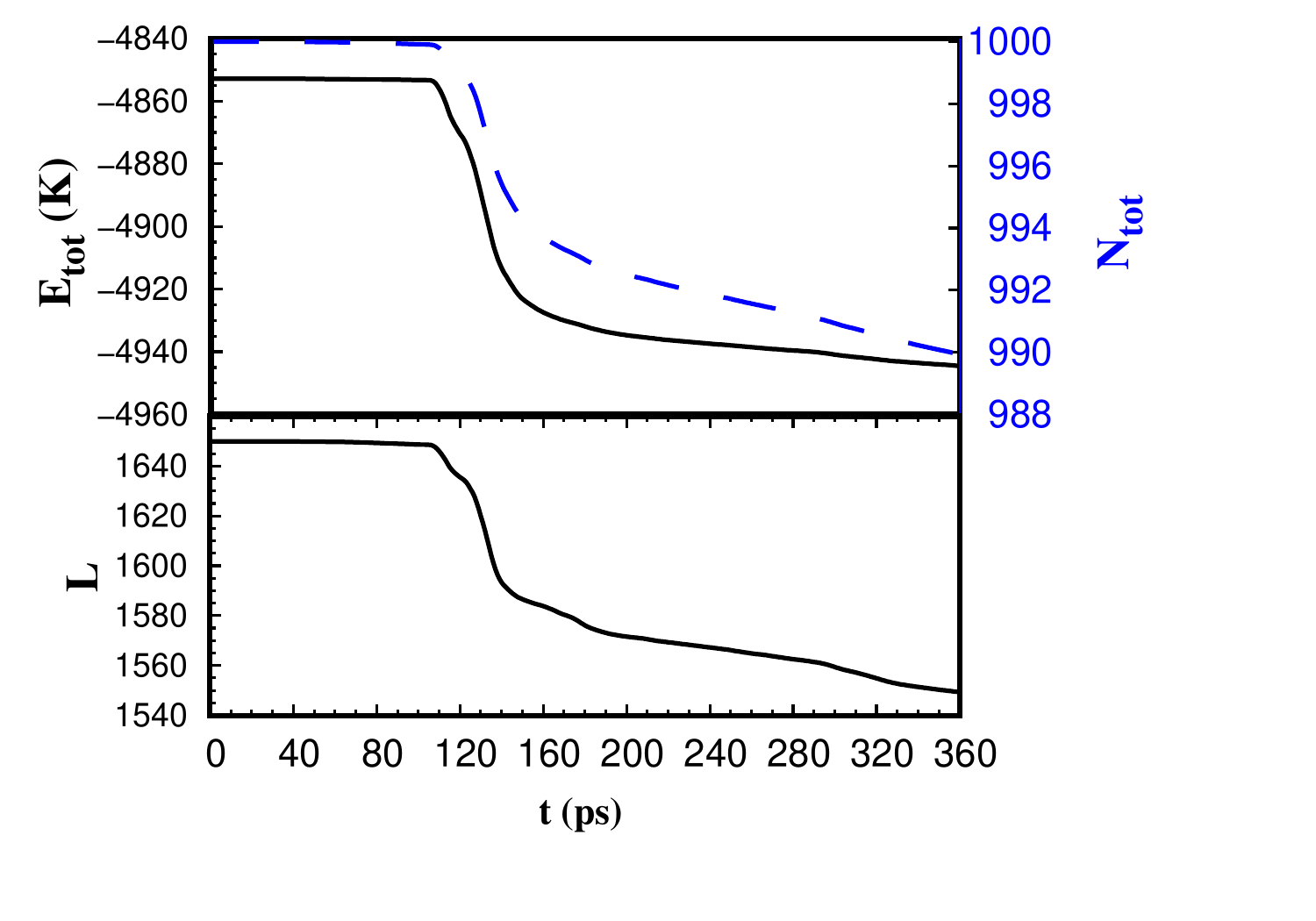}
\caption{\label{fig:b3R2_v02-EN-L}
Bottom plot: time evolution of the total angular momentum for He$_{500}$ + He$_{500}$ collision with impact parameter $b=3R/2$ and $v=20$~m/s. 
Top plot: Time evolution of the total energy $E_\text{tot}$ (black line, left vertical axis) and number of helium atoms $N_\text{tot}$ (dashed blue line, right vertical axis)   for the same collision.}
\end{figure}

The angular momentum involved in these collisions is $L=1650$ and 3300, respectively.
The collision dynamics is similar to the case with $v=10$~m/s, see the corresponding movies in the supplementary material. 
Vortices are nucleated  by the same mechanism at indentations appearing on the fused droplet surface.
We have found that 
the number of vortices of equal circulation increases from two at 10 m/s, to four at 20 and 40 m/s. During the real time evolution 
of the fused droplet,  some of these vortices are  evaporated.
This does not mean that $L$ changes, angular momentum simply goes into capillary waves. 
The interplay  between vortices and capillary waves is readily seen in these movies, which also show the tendency of increasing the
number of stable nucleated vortices with increasing droplet velocity. 

The time evolution of the total angular momentum  for $v=20$ m/s  is displayed in Fig.~\ref{fig:b3R2_v02-EN-L},
in addition to that of the total energy and number of atoms in the merged droplet.
As expected, the decrease in angular momentum follows that in energy and it is due to helium 
atoms evaporating from the droplet which are removed from the simulation box by the action of the absorbing buffer. 
During the 360 ps covered by the simulations, about 10 He atoms are evaporated for the collision at $v=20$ m/s taking away an
energy of 9.2 K/atom and an angular momentum of 10 units per atom. At $v=40$ m/s, we have found that 
15 He atoms are evaporated, taking away an average energy of  6.3 K/atom and an angular momentum of 16 units per atom.

\subsection{Collisions with impact parameter $b=2R$}
Grazing collisions are especially relevant since it is not obvious that the vdW attraction between the colliding droplets may compensate 
the kinetic energy in the colliding droplets and lead to droplet coalescence.

\subsubsection{$v=$ 20 m/s and  40 m/s collisions } 
 
\begin{figure}[t]
 \hspace*{-0.5cm}
\includegraphics[width=1.1\linewidth,clip=true]{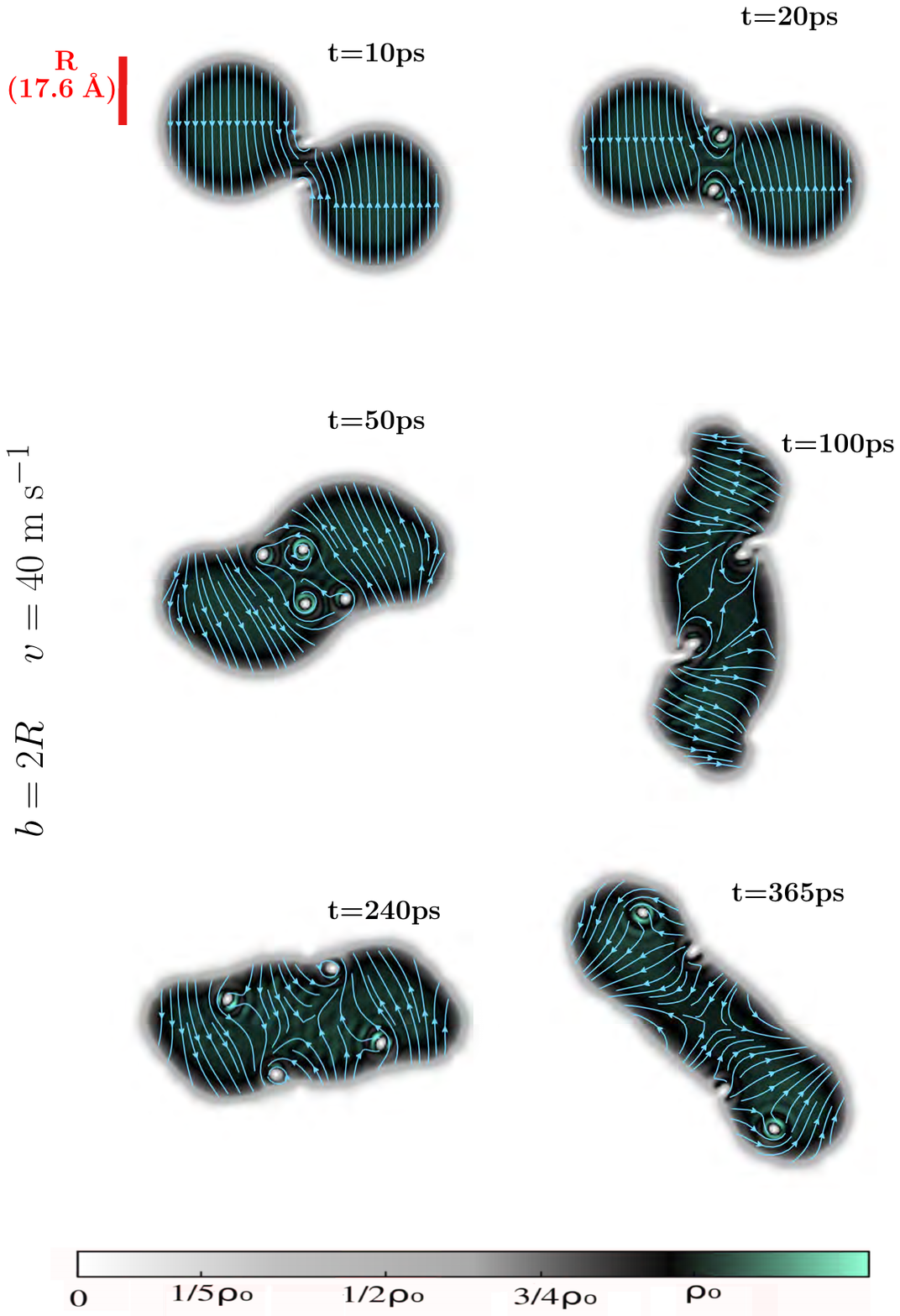}
\caption{\label{fig:b2R-v04-multiplots} 
Snapshots of 2-dimension cuts in the $(y,z)$ collision plane for impact parameter $b=2R$ (``grazing collision'') and $v=40$~m/s.\\
}
\end{figure}

The angular momentum involved in these distal collisions is $L=2200$ and 4400, respectively.
Figure~\ref{fig:b2R-v04-multiplots} 
 shows snapshots of the 2D density for the $v=40$ m/s case.
A density bridge perpendicular to the collision direction appears, connecting both droplets (frame at 10 ps). A vortex dimer is
nucleated at $t=20$ ps, and another dimer appears at 50 ps. The interplay between capillary waves and vortices leads to the evaporation 
of one of the vortex dimers (frame at 110 ps) which is nucleated again later on (frame a 240 ps) and re-evaporated at $t=365$ ps.
The coalesced droplet is very stretched due to the angular momentum deposited in the system. 

During the 475 ps elapsed by the real time simulations,  about 5 He atoms are evaporated, taking away an
energy of 13.6 K/atom and an angular momentum of 23 units per atom. 
As shown in the movies, the evolutions at $v=20$ and 40 m/s are qualitatively similar.
 
\subsection{Distal collision at $b=5R/2$ and $v=$ 20 m/s}   
 
\begin{figure}[t]
 \hspace*{-0.5cm}
\includegraphics[width=1.1\linewidth,clip=true]{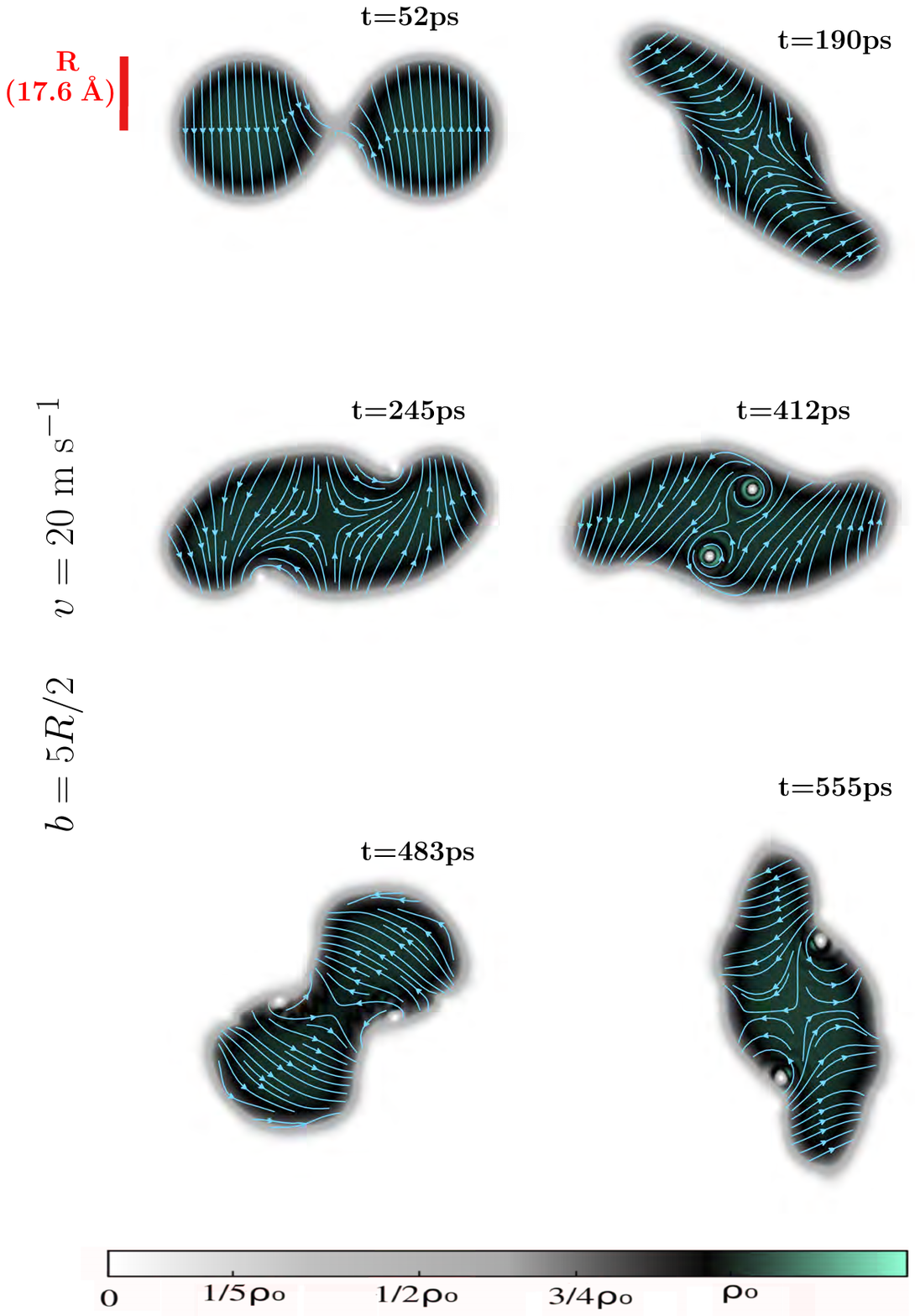}
\caption{\label{fig:b5R2-v02-multiplots} 
Snapshots of 2-dimension cuts in the $(y,z)$ collision plane for impact parameter $b=5R/2$ (``distal collision'') and $v=20$~m/s.\\
}
\end{figure}

The angular momentum involved in this distal collision is $L=2750$.
This collision highlights the relevance of the finite range of the vdW interaction in the outcome of the process. Indeed, if the collision were
modelled by  a surface tension plus kinetic energy model, inherent to any classical model based on the NS approach, it would lead to the non-interaction 
of the approaching droplets. At variance, we have found that the colliding droplets merge. 

Figure~\ref{fig:b5R2-v02-multiplots} 
shows snapshots of the 2D density.
A tiny, low-density bridge is clearly visible at 52 ps. Eventually, droplets merge yielding a vortex-free droplet for a relatively long amount of time, as
illustrated by the frame at $t=$ 190 ps, where the merged droplet  undergoes a complete rotation 
with all the angular momentum stored in the form of capillary waves. Eventually,
a vortex dimer starts being nucleated at $t=$ 245 ps by the familiar surface indentations mechanism; it is
clearly visible e.g. at $t=$ 412 ps. The vortex dimer later evaporates (frame at $t=$ 483 ps) but it is nucleated again at $t=$ 555 ps. This 
evaporation-nucleation process continues until the end of the real time simulation
(691 ps). During the  
time elapsed by the  simulation,  about 3 He atoms are evaporated, taking away an
energy of 3.7 K/atom and an angular momentum of 22 units per atom. 
 
\subsection{Asymmetric collisions} 
We have seen that, due to the symmetry of the binary collision between two identical droplets, vortices are nucleated in pairs 
by the surface indentation mechanism.
A less symmetric collision might lead to the nucleation of an odd number of  vortices. To check this possibility, 
and see the influence of the asymmetry on the collision outcome, we have 
conducted one simulation with droplets of different sizes, namely $N_1=300$ and $N_2=700$. The initial conditions 
  $v_1=28$ m/s, $v_2= 12$ m/s, $b = (25/21)(3R/2)$ were 
 chosen so as to be as close as possible to the case of identical droplets with $v = 20$ m/s and $b = 3R/2$, in order to compare the 
 collision processes for the same relative velocity and total angular momentum.

\begin{figure}[t]
 \hspace*{-0.5cm}
\includegraphics[width=1.1\linewidth,clip=true]{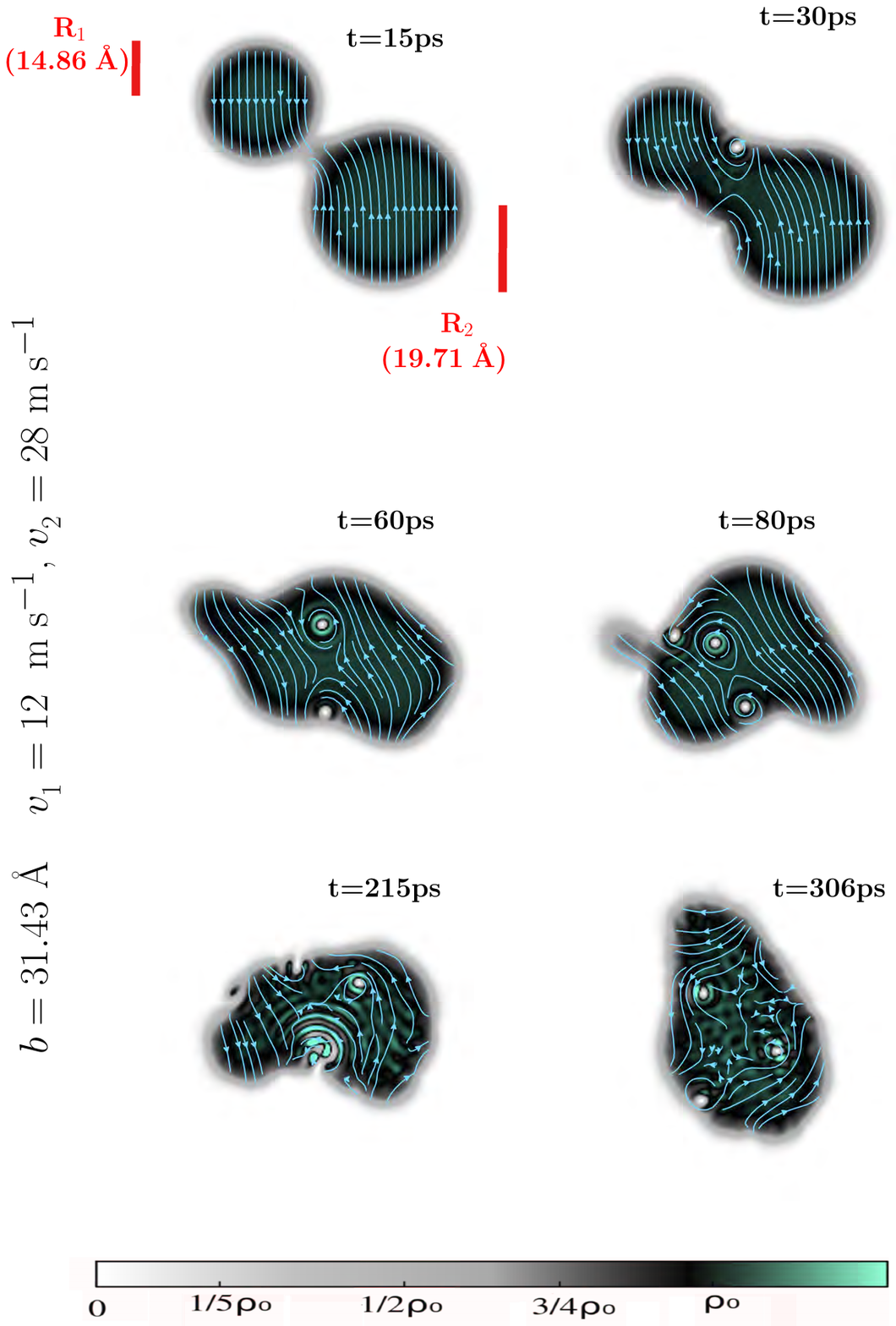}
\caption{\label{fig:He300-He700-multiplots} 
Snapshots of 2-dimension cuts in the $(y,z)$ collision plane for He$_{300}$+He$_{700}$ asymmetric collision with impact parameter $b=31.4$ \AA{} and velocities $v_1=28$~m/s (upper left droplet, $N_1=300$) and $v_2=12$~m/s (lower right droplet, $N_2=700$), chosen to be as close as possible to the $b=3R/2$, $v=20$~m/s case for two identical, $N=500$ droplets (same relative velocity and total angular momentum).}
\end{figure}

Figure~\ref{fig:He300-He700-multiplots} 
 shows snapshots of the 2D density.
The usual density bridge can be seen at 15 ps, and one single vortex is nucleated at 30 ps. Yet, another vortex is later nucleated at 60 ps, and a third one appears 
at 80 ps. The latter panel also shows a surface protrusion whose collapse yields a vortex ring and a series of density waves
propagating inside the droplet. 
One of the vortices gets ejected at $t=215$~ps, but then it gets nucleated again as can be seen in the  final snapshot at $t=306$~ps.
During the whole simulation,  about 7 He atoms are evaporated, taking away an
energy of $8.6$~K/atom and an angular momentum of 16.1 units per atom. 

\subsection{Sharing angular momentum between capillary waves and vortex lines}\label{ssec:cap-vx-L}
It is well known that angular momentum in superfluid $^4$He droplets can be stored in the form of capillary waves and/or quantized vortices,
see e.g. Refs. \onlinecite{Anc18,Oco20,Pi21}.
As discussed in the previous Sections, and as it is clearly apparent from the figures,
both vortices and  capillary waves appear in the merged droplets.
It is quite natural to ask oneself how much angular momentum is stored into vortices and how much is in capillary waves.
This question has not a rigorous answer, as one cannot split the effective wave function of the superfluid $\Psi(\mathbf{r}, t)$ into a component
arising from vortex contributions and another one from capillary waves, both being intimately entangled.

A simple estimate of vortex ($L_v$) and capillary wave ($L_{cap}$) contributions to the total angular momentum can be obtained as done in Refs.
\onlinecite{Oco20,Pi21}, when the shape of the rotating droplet is approximately ellipsoidal. It consists in determining $L_{cap}$ from the angular
velocity $\omega$ of the apparent rotation of the merged droplet and using an ellipsoid approximation for the droplet shape, since
 the angular momentum of an ellipsoid made of an irrotational fluid rotating around a principal axis at angular velocity $\omega$ is known.\cite{Cop17b} $L_v$ is  obtained as $L-L_{cap}$.
These are only estimates that could be more meaningful near the end of the simulations when the droplet reaches a quasi 
steady rotational state.

We have proceeded as follows. We  first determine  
the classical axes of inertia  by diagonalizing the classical matrix of inertia in the lab frame,
\begin{equation}
I_{jk}=m_4 \int d\mathbf{r}\, (r^2 \delta_{jk}- r_j r_k) \, \rho(\mathbf{r}) \, .
\label{eq8}
\end{equation}
Since the $x$ axis is maintained constant by symmetry, the instantaneous inertia axes were determined by rotation by a single 
angle $\theta$ about $x$. The angular velocity $\omega$ is then calculated as
\begin{equation}
\omega=\frac{\Delta \theta}{\Delta t} \, .
\label{eq9}
\end{equation}
The angular momentum due to capillary waves is finally expressed as $L_{cap}=\mathcal{I}_\mathrm{irr} \omega$, 
where\cite{Cop17b}
\begin{equation}
\mathcal{I}_\mathrm{irr}=
	m_4 \,N_\text{tot}~\frac{[\langle y^2 \rangle -\langle z^2 \rangle]^2}{\langle y^2 \rangle+\langle z^2\rangle}
\label{eq10}
\end{equation}
is  the irrotational moment of inertia calculated in the rotating frame, with
\begin{equation}
	\langle y^2\rangle=\frac{1}{N_\text{tot}}\int d\textbf{r}~y^2~\rho(\textbf{r})
	\quad\text{and}\quad
   \langle z^2 \rangle=\frac{1}{N_\text{tot}}\int d\textbf{r}~z^2~\rho(\textbf{r}) \, ,
   \label{eq11}
\end{equation}
$N_{tot}$ being the total number of atoms in the merged droplet.
For vortex-free droplets, the above expressions have been found to reproduce the DFT results within $5\%$.\cite{Pi21}
 
\begin{figure}[t]
 \hspace*{-0.3cm}
\includegraphics[width=0.74\linewidth,angle=270,clip=true]{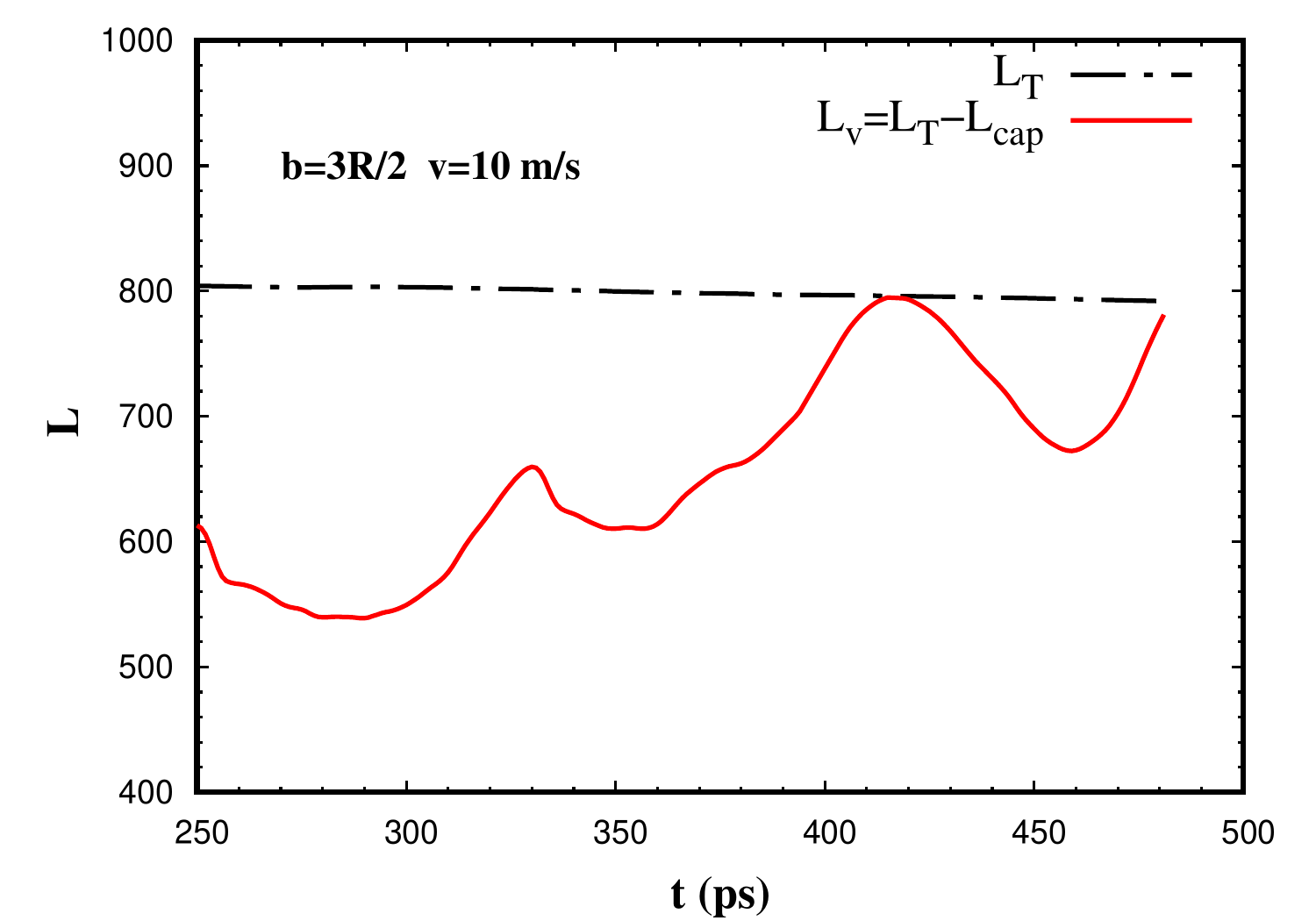}
\caption{\label{fig:b3R2-v01-Lvortex} 
Time evolution of the total angular momentum and of the vortex contribution $L_{\mathrm{v}}$ estimated by subtracting the angular momentum due to capillary waves $L_\mathrm{cap}$ from the total angular momentum $L$ (see text), for He$_{500}$+He$_{500}$ merging collision with impact parameter $b=3R/2$ and initial impinging velocity $v=10$~m/s.\\
}
\end{figure}

\begin{figure}[t]
 \hspace*{-0.3cm}
\includegraphics[width=0.74\linewidth,angle=270,clip=true]{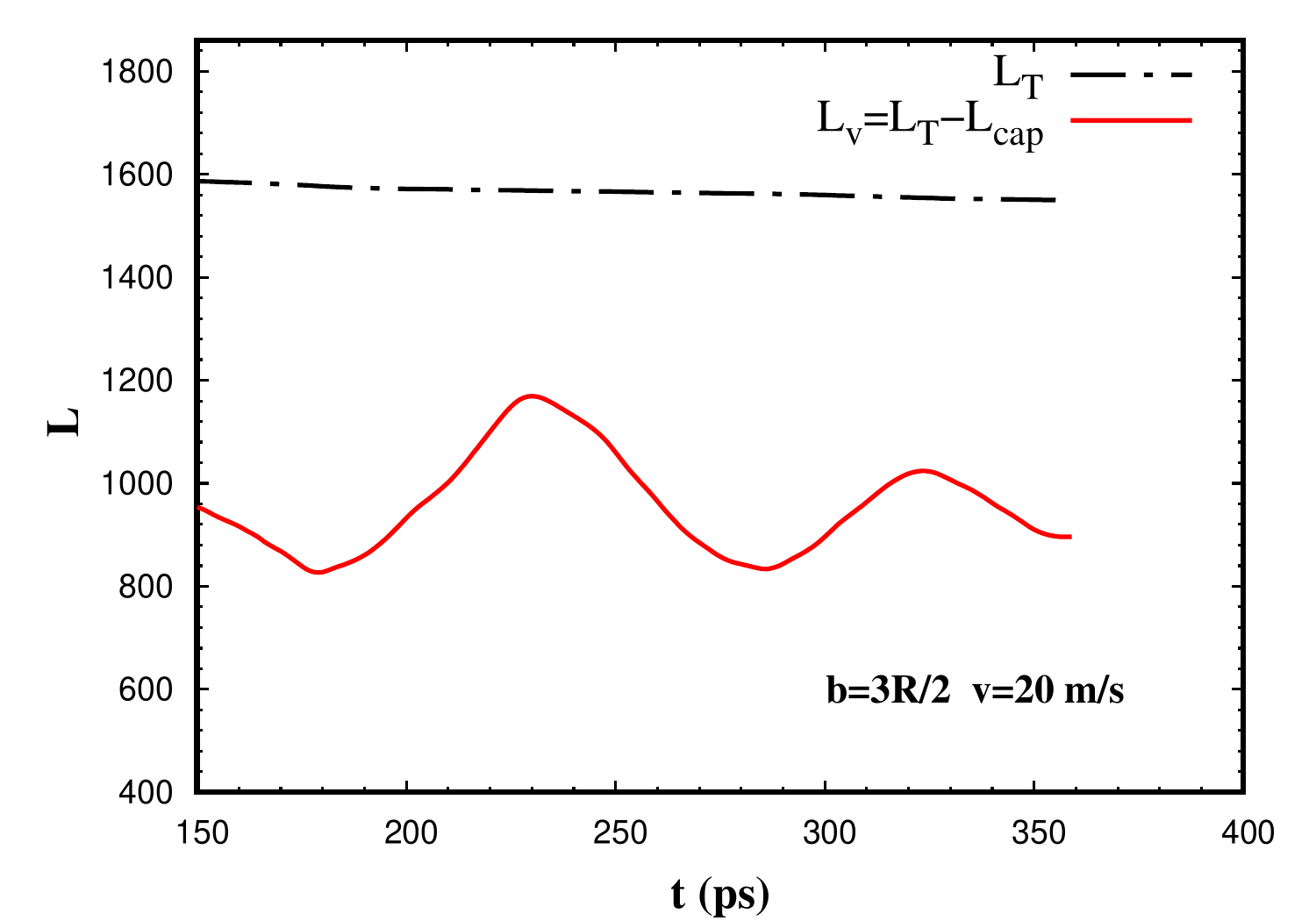}
\caption{\label{fig:b3R2-v02-Lvortex} 
Time evolution of the total angular momentum and of the vortex contribution $L_{\mathrm{v}}$ estimated by subtracting the angular momentum due to capillary waves $L_\mathrm{cap}$ from the total angular momentum $L$ (see text), for He$_{500}$+He$_{500}$ merging collision with impact parameter $b=3R/2$ and initial impinging velocity $v=20$~m/s.}
\end{figure}

As illustrative examples, we show in 
Fig.~\ref{fig:b3R2-v01-Lvortex} 
 the total angular momentum $L$ (which changes with time due to atom evaporation)
and vortex contribution $L_v$ for the collision corresponding to $b=3R/2$ and $v=$ 10 m/s,
 for  times $t >$ 250 ps. 
Figure~\ref{fig:b3R2-v02-Lvortex} 
shows the same quantities for the collision corresponding to $b=3R/2$ and $v=$ 20 m/s, and times $t >$ 150 ps.
We want to stress here that these qualitative results should be taken with caution, 
as some of the considered droplet configurations are not 
as ellipsoidal as they should be to justify the application of the above expressions.

\section{Summary and concluding remarks}

We have addressed  binary collisions of superfluid helium drops within the  He-DFT approach.
The simulations have been carried out  for $^4$He$_{500}$ droplets and several values of the impact parameter
and relative velocities. To see the influence of droplet asymmetry, we have also addressed the collision of two droplets with different number of atoms. Asymmetric
collisions seem to favor the appearance of an odd number of vortices, whereas this number can only be even in binary collisions of equal size droplets.

Not surprisingly, collisions of superfluid $^4$He droplets  display similarities with classical droplet collisions. In both cases, the merged droplet is 
highly deformed and rotates in order to maintain the angular momentum 
involved in the collision; compare, e.g., the morphology of the 
droplets shown in Ref. \onlinecite{Nik09} with those displayed in this work.
The substantial difference between both situations is the ubiquitous appearance of quantized vortices in the case 
of helium, made possible within the He-TDDFT framework. 
Besides this important point, the  He-TDDFT approach differs from 
classical ones based on the 
solution of the Navier-Stokes or Euler equations
in which the former 
takes into account the finite range of the van der Waals 
interaction which facilitates droplet merging for grazing and distal collisions, whereas it is not
possible for the latter approaches where droplet interaction is mediated by surface tension and kinetic energy of the colliding droplets.  

Computational limitations make it impossible to implement the He-TDDFT method in the experimental conditions under which very large droplets are made, which involve much larger number of atoms at smaller
relative velocities.\cite{Tan20,Kol22,Ulm23}  Yet, an interesting conclusion is readily transferable to that experimental situation.
Using a nozzle shape specifically devised to reduce angular momentum acquisition when droplets travel through the source chamber, 
a recent experiment\cite{Ulm23} still identified a few vortex-hosting droplets from the appearance of Xe filament-shaped structures in 
x-ray diffraction images.
 These observations suggest that droplet collisions produced during the
expansion from the source chamber might be the cause of angular momentum acquisition and subsequent vortex nucleation. 
Our calculations make this scenario plausible.
On the one hand, we have found that quantum vortices are readily nucleated by the surface indentations mechanism, yielding vortex rings
(which carry no angular momentum)  for
head-on collisions, and off center vortices (which carry angular momentum, although smaller that centered vortices) for non-zero impact parameter collisions.
Since indentations appear whenever droplets merge, the indentation mechanism is independent on the droplet size.

In addition, we have unexpectedly found that, even for grazing and distal collisions, droplets coalesce
at  relative velocities as large as 40 m/s instead of stretching and separating again; these velocities are 
much larger than those found in the experiments.\cite{Tan20,Kol22,Ulm23}  
Thus, droplet collisions in a broad interval of impact parameters and relative velocities would lead to vortex nucleation.

Our simulations also show that droplet-droplet collisions could also nucleate vortices in smaller droplets, in the range of a thousand atoms.
Their appearance would be favored in conditions where velocity spread is larger.
So far no convincing way of detecting them in small droplets has been demonstrated.

He-TDDFT simulations have other unavoidable limitations.
On the one hand, the method  is strictly a zero temperature approach and there is no dissipation; energy can only be lost by atom evaporation, whereas 
any residual viscosity remaining in the system would contribute to stabilize the merged droplet and damp density oscillations.
On the other hand, due to the 
limited time elapsed by the simulations, droplets do not reach the stationary state of apparent rotation and stabilized vortex array structures found in the
experiments.\cite{Gom14,Oco20} 

\section*{SUPPLEMENTARY MATERIAL}   
See supplementary material for the video files showing the real time evolution
of the processes discussed in the present work.

 \begin{acknowledgments}
 We are very indebted to Rico Tanyag and Thomas M\"oller for useful exchanges.
A computer grant from  CALMIP high performance computer center (grant P1039) is gratefully acknowledged.
This work has been  performed under Grant No.  PID2020-114626GB-I00 from the MICIN/AEI/10.13039/501100011033
and benefitted from COST Action CA21101 ``Confined molecular systems: 
form a new generation of materials to the stars'' (COSY) 
supported by COST (European Cooperation in Science and Technology).
\end{acknowledgments}

\bigskip

\section*{AUTHOR DECLARATIONS}   
\subsection*{Conflict of Interest}
The authors have no conflicts to disclose.


\subsection*{DATA AVAILABILITY}
The data that support the findings of this study are available
from the corresponding author upon reasonable request

\end{document}